\documentclass[a4paper,12pt]{article}

\usepackage{amssymb, amsmath}
\usepackage{hyperref}

\newcommand{\ket}[1]{| #1 \rangle}
\newcommand{\bra}[1]{\langle #1 |}
\newcommand{\inprod}[2]{\langle #1 | #2 \rangle}
\newcommand{\oper}[2]{| #1 \rangle \langle #2 |}

\DeclareMathOperator{\Tr}{Tr}

\newtheorem{thm}{Theorem}
\newtheorem{lemma}{Lemma}

\begin{document}

\title{Constructions of indecomposable positive maps based on a new criterion for indecomposability}
\author{William Hall \\ \small Department of Mathematics, University of York \\ \small Heslington, York YO10 5DD, United Kingdom \\ \small \texttt{wah500@york.ac.uk}  }

\maketitle
\begin{abstract}
We give a criterion for a positive mapping on the space of operators on a Hilbert space to be indecomposable. We show that this criterion can be applied to two families of positive maps. These families of maps can then be used to form separability criteria for bipartite quantum states that can detect the entanglement of \emph{bound entangled} quantum states.
\end{abstract}

\section{Introduction}

The phenomenon of entanglement \cite{Ent} in quantum mechanics plays a central role in quantum information theory and quantum computation \cite{Ent2, Ent3}. However, for a general bipartite mixed quantum state, it is still not known how to determine whether such a state is entangled or separable. Since the results of many experiments are not pure but in fact mixed states, this is a problem that is of both fundamental and practical importance within quantum mechanics and quantum information theory.

One of the most studied mathematical tools for determining whether a quantum state is entangled or not is the theory of positive maps of operators on a Hilbert space \cite{H1}. In particular, the construction of \emph{indecomposable positive maps} is a topic of particular importance, because they can be used to form strong necessary criteria for a quantum state to be separable. Athough much study has been dedicated to this topic, only specific examples of these maps have been found \cite{Ch2, ChGen, ChGen2, ChGen3, ChGen4, ChGen5, ChGen6, Piani, UPBmap, Rob1, Rob2, sg, K1, K2, ptn2}, and there are no general constructions of nondecomposable maps.

St{\o}rmer gave a necessary and sufficient condition for a positive map to be decomposable \cite{St_dec}; however, in practice, this condition is very hard to verify. What we give here is a \emph{sufficient condition for a positive linear map to be indecomposable}. Furthermore, this condition only requires the linear map to be expressible in a certain form, and so is easier to use in practice.

This paper proceeds as follows. In Section \ref{sec_ent}, we recall some of the known facts about entanglement of quantum states and the role of positive maps in showing quantum states are entangled. In Section \ref{sec_crit}, we state and prove our general criterion for a positive map to be indecomposable. In Section \ref{sec_ex}, we give two families of maps that can be shown to be indecomposable using this criterion. We conclude with a discussion in section \ref{sec_con}.

\section{Entanglement in bipartite systems and positive maps} \label{sec_ent}

\subsection{Entangled and separable quantum states}

We will begin by recalling some basic facts about entanglement of bipartite quantum systems. Let $\mathcal{H}_i \ (i=1,2)$ be finite dimensional Hilbert spaces. We define $\mathcal{B}(\mathcal{H})$ to be the set of (bounded) operators on a Hilbert space $\mathcal{H}$. A bipartite quantum state $\rho \in \mathcal{B}(\mathcal{H}_1 \otimes \mathcal{H}_2)$ is said to be \emph{separable} \cite{Wsep} if and only if $\rho$ can be written in the form
\begin{equation} \rho = \sum_{i=1}^k p_i \rho^{(1)}_i \otimes \rho^{(2)}_i \label{sep} \end{equation}
where $p_i > 0, \sum_i p_i = 1$ and $\rho^{(j)}_i \in \mathcal{B}(\mathcal{H}_j)$ are density matrices on the individual parts of the bipartite system. A bipartite quantum state is said to be \emph{entangled} if it is not separable.

\subsection{Positive maps and entanglement}

We will also recall some basic facts about positive maps, and how they can be used to show that a bipartite quantum state is entangled \cite{H1}. Let $\Lambda: \mathcal{B}(\mathcal{H}) \to \mathcal{B}(\mathcal{H})$ be a linear map. $\Lambda$ is said to be a \emph{positive map} if, for all Hermitian $\sigma \in \mathcal{H}$ such that $\sigma \geq 0$, then $\Lambda(\sigma)$ is Hermitian and $\Lambda(\sigma) \geq 0$. A stronger condition on $\Lambda$ is that of complete positivity: $\Lambda$ is \emph{completely positive} if, for all Hilbert spaces $\mathcal{K}$, the map $I \otimes \Lambda: \mathcal{B}(\mathcal{K} \otimes \mathcal{H}) \to \mathcal{B}(\mathcal{K} \otimes \mathcal{H})$ (where $I$ is the identity map) is also positive. We say $\Lambda$ is \emph{completely co-positive} if $\Lambda \circ T$ is completely positive, where $T$ is the transpose map. Due to work by Choi \cite{Ch1} and Kraus \cite{K}, it has been shown that a completely positive map $\Lambda: \mathcal{B}(\mathcal{H}) \to \mathcal{B}(\mathcal{H})$ has a decomposition of the form $\Lambda(\rho) = \sum_i V_i\rho V_i^\dagger$. The operators $V_i \in \mathcal{B}(\mathcal{H})$ are known as \emph{Kraus operators}.

The existence of positive maps that are not completely positive allows us to use these maps to produce \emph{sufficient} conditions for a bipartite quantum state $\rho \in \mathcal{B}(\mathcal{H}_1 \otimes \mathcal{H}_2)$ to be separable i.e. for $\rho$ to have a decomposition of the form of (\ref{sep}). Then, for any positive map $\Lambda: \mathcal{B}(\mathcal{H}_2) \to \mathcal{B}(\mathcal{H}_2)$,
\begin{equation} (I \otimes \Lambda)(\rho) = \sum_i p_i \rho^{(1)}_i \otimes \Lambda(\rho^{(2)}_i) \geq 0. \end{equation}
Hence if $(I \otimes \Lambda)(\rho) \ngeq 0$, the bipartite state $\rho$ must be entangled. We say that this test \emph{detects} the entanglement of $\rho$.

An example of a positive linear map that is not completely positive is the transpose map. This leads to the necessary \emph{partial tranpose condition} for separability \cite{pt}, that $(I \otimes T)(\rho)\geq 0$. For a $2 \otimes 2$ or $2 \otimes 3$ bipartite system\footnote{The notation $d_1 \otimes d_2$ for $d_1, d_2 \in \mathbb{N}$ denotes the dimensions of the individual Hilbert spaces.}, this condition is sufficient as well \cite{H1, ptn1, ptn2}. More generally, there exist entangled states that do not violate this condition. These states are known as \emph{bound entangled} states and play an important role in the theory of quantum information \cite{bes}.

The partial tranpose condition is computationally a very simple condition to verify, and so it is important to know when a condition for separability produced from a positive map does not detect any states that the partial transpose condition does not detect. A positive map $\Lambda: \mathcal{B}(\mathcal{H}) \to \mathcal{B}(\mathcal{H})$ is said to be \emph{decomposable} if it can written as the sum of a completely positive and a completely co-positive map. For such maps, by the definition of completely positive and completely co-positive,
\begin{equation} (I \otimes T)(\rho) \geq 0 \Rightarrow (I \otimes \Lambda)(\rho) \geq 0 \end{equation}
i.e. the partial transpose condition detects any entanglement that the condition formed from $\Lambda$ does. A positive map is called \emph{indecomposable} if such a decomposition for $\Lambda$ does not exist. 

Separability criteria formed from indecomposable maps are hence important because they are the only criteria formed from positive maps that can be used to detect bound entangled quantum states, and a number of examples exist in the literature \cite{Ch2, ChGen, ChGen2, ChGen3, ChGen4, ChGen5, ChGen6, Piani, UPBmap, Rob1, Rob2, sg, K1, K2, ptn2}.

St{\o}rmer gave the following necessary and sufficient condition for a linear map to be positive:

\begin{thm} Let $\Lambda: \mathcal{B}(\mathcal{H}) \to \mathcal{B}(\mathcal{H})$ be a linear map. Then $\Lambda$ is decomposable if and only if, for all $n \in \mathbb{N}$, and all positive operators $\rho \in \mathcal{B}(\mathbb{C}^n \otimes \mathbb{H})$ such that $(T \otimes I)(\rho)$ is also positive, $(I \otimes \Lambda)(\rho)$ is also positive. \end{thm}

Since we do not know how to completely characterise the positive operators given in the statement of the theorem, in practice we cannot use this theorem to show a map is decomposable. However, in some cases, the theorem has been used to show a positive map is not decomposable, by constructing such a positive operator $\rho$ such that $(I \otimes \Lambda) \ngeq 0$. The condition we will give in section \ref{sec_crit} will not require the construction of such an operator.

\subsection{Entanglement witnesses}

We conclude our review of positive maps by introducing the notion of an \emph{entanglement witness}, and studying their relation to positive maps. An operator $W \in \mathcal{B}(\mathcal{H}_1 \otimes \mathcal{H}_2)$ is an entanglement witness if $\bra{\psi}W\ket{\psi} \geq 0$ for all $\ket{\psi} = \ket{\psi_1}\ket{\psi_2}$. Hence if a bipartite state $\rho \in \mathcal{B}(\mathcal{H}_1 \otimes \mathcal{H}_2)$ is such that $\Tr (W \rho) < 0$ for some entanglement witness $W$, $\rho$ must be entangled.

Choi \cite{Ch2} and Jamio{\l}kowski \cite{J} independently discovered that there is a one-to-one correspondence between positive maps and entanglement witnesses. This can be seen in the following way: Let $\mathcal{H}$ be a finite dimensional Hilbert space with orthonormal basis $\{ \ket{k} \}$. For any linear map $\Lambda: \mathcal{B}(\mathcal{H}) \to \mathcal{B}(\mathcal{H})$, we can define a 4-index array $\Lambda_{ijkl}$ by
\begin{equation} \Lambda(\oper{k}{l}) = \sum_{k,l} \Lambda_{ijkl} \oper{i}{j}.\end{equation}
This 4-index array is essentially what defines an entanglement witness when $\Lambda$ is positive. The Jamio{\l}kowski form of this correspondence \cite{J} can be written in the form
\begin{equation} W_\Lambda = (\Lambda \otimes I) \left( \sum_{k,l} \oper{k}{l} \otimes \oper{k}{l} \right) = \sum_{ijkl} \Lambda_{ijkl} \oper{i}{j} \otimes \oper{k}{l} \end{equation} 
which when $\Lambda$ is positive defines an entanglement witness $W_\Lambda \in \mathcal{B}(\mathcal{H} \otimes \mathcal{H})$ \cite{J}.  Furthermore,
\begin{enumerate}
\item If $\Lambda(\rho) = \sum_a V_a \rho V_a^\dagger$ (completely positive), then the associated entanglement witness is 
\begin{equation} W_\Lambda = \sum_a \sum_{ijkl} (V_a)_{ik}(V_a)^*_{jl} \oper{i}{j} \otimes \oper{k}{l} = \sum_a \oper{V_a}{V_a} \label{pf_cp} \end{equation}
where if $V \in \mathcal{B}(\mathcal{H})$, then $\ket{V} = \sum_{i,j} V_{ij}\ket{i}\ket{j} \in \mathcal{H} \otimes \mathcal{H}$;
\item If $\Lambda(\rho) = \sum_a V_a\rho^T V_a^\dagger$ (completely copositive), then the associated entanglement witness is
\begin{equation} W_\Lambda = \sum_a \sum_{ijkl} (V_a)_{il}(V_a)^*_{jk} \oper{i}{j} \otimes \oper{k}{l} = \sum_a \oper{V_a}{V_a}^{T_B} \end{equation}
where $T_B$ represents partial transposition of the second system in the tensor product $\mathcal{H} \otimes \mathcal{H}$ with respect to the chosen basis.
\end{enumerate}

All of these observations together give us the following result (also proved in e.g. \cite{Red1}:

\begin{lemma} If $\Lambda : \mathcal{B}(\mathcal{H}) \to \mathcal{B}(\mathcal{H})$ is a decomposable positive map, then the associated entanglement witness $W_\Lambda$ can be expressed in the form 
\begin{equation} W_\Lambda = P + Q^{T_B}  \label{d} \end{equation} 
where $P,Q$ are positive operators.
\end{lemma}

\section{A criterion for indecomposability} \label{sec_crit}

\subsection{The general idea}

When given the entanglement witness $W_\Lambda$ associated with a positive map $\Lambda$, the difficulty in trying to show whether we can write $W_\Lambda$ in the form of (\ref{d}) is that such a decomposition may not be unique, and certain terms in $W_\Lambda$ may arise from either the $P$ or the $Q^{T_B}$ term. Two examples are:
\begin{enumerate}
\item Separable terms e.g.\footnote{The stars in the below expression denote complex conjugation in the standard basis i.e. if $\ket{x}=\sum x_k\ket{k}$, then $\ket{x^*}= \sum x_k^* \ket{k}$.}
\begin{equation} \oper{a}{a} \otimes \oper{b}{b} = \oper{a}{a} \otimes \oper{b^*}{b^*}^T = \left( \oper{a}{a} \otimes \oper{b^*}{b^*} \right)^{T_B}; \end{equation}
\item 'Bound entangled' terms i.e. if $P \in \mathcal{B}(\mathcal{H} \otimes \mathcal{H})$ is entangled and $P^{T_B}$ is positive, then $P$ = $\left( P^{T_B} \right)^{T_B}$.
\end{enumerate}
More generally, if $P$ (or similarly $Q$) contains terms that are positive under a partial transpose, then such terms can arise from either term in the decomposition (\ref{d}). 

What we aim to do here is to present conditions on the positive map $\Lambda$ that mean this ambiguity cannot arise, and if we can write $W_\Lambda$ in the form of (\ref{d}), then we will know which terms should arise from the $P$ term and those which must arise from the $Q^{T_B}$ term. The idea will then be to show that under further conditions $P$ or $Q$ cannot be positive, contradicting the form of the expression of (\ref{d}), and hence forcing $\Lambda$ to be indecomposable.

\subsection{Statement and proof of criterion}

Let us set up some preliminaries. Let $\mathcal{V}$ be a linear subspace of $\mathcal{B}(\mathcal{H})$ of dimension $N$, and from this let us define the subspace $\mathcal{W}(\mathcal{V}) \subset \mathcal{H} \otimes \mathcal{H}$ by
\begin{equation} 
\mathcal{W}(\mathcal{V}) = \{ \ket{V} \ | \ V \in \mathcal{V} \}
\end{equation}
and furthermore we denote the subspace orthogonal to $\mathcal{W}(\mathcal{V})$ by $\mathcal{W}(\mathcal{V})^\perp$ i.e.
\begin{equation}
\mathcal{W}(\mathcal{V})^\perp = \{ \ket{\psi} \in \mathcal{H} \otimes \mathcal{H}  \ | \ \inprod{V}{\psi}=0 \ \forall \ \ket{V} \in \mathcal{W}(\mathcal{V}) \}.
\end{equation}

With these ideas in place, we are ready to state the main result of this paper.

\begin{thm} \label{id_test} Suppose $\Lambda: \mathcal{B}(\mathcal{H}) \to \mathcal{B}(\mathcal{H})$ is a positive map of the form
\begin{equation} \Lambda(\rho) = \sum_{m,n=1}^N \lambda_{mn} V_m\rho V_n^\dagger \end{equation} 
where the matrix $L$ defined by $\lambda_{mn}$ is Hermitian, and the set $\{V_k \}_{k=1}^N$ forms a basis for a subspace $\mathcal{V} \subset \mathcal{B}(\mathcal{H})$ such that the subspace $\mathcal{W}(\mathcal{V}) \subset \mathcal{H} \otimes \mathcal{H}$ defined as above has the property that for all $Q \in \mathcal{B}(\mathcal{H})$ positive, there exists $\ket{\psi} \in \mathcal{W}(\mathcal{V})^{\perp}$ such that $\bra{\psi}Q^{T_B}\ket{\psi}>0$. Then if $L$ has a negative eigenvalue, then $\Lambda$ is not a decomposable map. \end{thm}

\textbf{Proof} The entanglement witness $W_\Lambda$ associated with $\Lambda$ is given by 
\begin{equation} W_\Lambda = \sum_{m,n=1}^N \lambda_{mn} \oper{V_m}{V_n}. \label{f1} \end{equation}
Now suppose $\Lambda$ is decomposable i.e. $W_\Lambda=P+Q^{T_B}$ with $P,Q>0$. By the definition of $\mathcal{W}(\mathcal{V})$, $W_\Lambda \in \mathcal{B}(\mathcal{W}(\mathcal{V}))$. Furthermore, we can write $P = W - Q^{T_B}$. However, by hypothesis, there exists $\ket{\psi} \in \mathcal{W}(\mathcal{V})^{\perp}$ such that $\bra{\psi}Q^{T_B}\ket{\psi}>0$, and hence
\begin{eqnarray}
\bra{\psi}P\ket{\psi} &=& \bra{\psi}W\ket{\psi} - \bra{\psi}Q^{T_B}\ket{\psi} \\
											&=& - \bra{\psi}Q^{T_B}\ket{\psi} < 0 
\end{eqnarray}
where the first expectation is zero because $\ket{\psi} \in \mathcal{W}(\mathcal{V})^\perp$ while $W_\Lambda \in \mathcal{B}(\mathcal{W}(\mathcal{V}))$. This contradicts the fact that $P$ is a positive operator, and so necessarily $Q=0$. Hence $W_\Lambda=P$ i.e. $W_\Lambda$ should be a positive operator. However, if the matrix $L$ defined from $\lambda_{mn}$ has a negative eigenvalue, so does $W_\Lambda$, giving us a contradiction. $\Box$.

To conclude this subsection, let us make a few observations about Theorem \ref{id_test}. First, Theorem \ref{id_test} is concerned with functions of $\rho$. However, all of these statements hold equally for functions of $\rho^T$ here, because it is clear from the definition of a decomposable map that if $\Lambda$ is a decomposable map, then so is the map $\Lambda \circ T$. In terms of the entanglement witness, the positive $P$ term would be restricted to be zero.

Secondly, it is worth noting that we did not simply assert that $\mathcal{W}(\mathcal{V})$ contained no operators with a positive partial transpose, but we asserted a condition which is at least as strong as this instead. This issue discussed in section \ref{sec_con}. Indeed, it is not even immediately obvious that any subspace $\mathcal{W}(\mathcal{V})$ can be constructed. We will see however from the examples in section \ref{sec_ex} that such subspaces do exist.

\subsection{Extensions of the criterion}

When we have a more general positive map of the form
\begin{equation} \Lambda(\rho) = \sum_{m,n=1}^N \lambda_{mn} U_m\rho U_n^\dagger + \sum_{p,q=1}^{N^\prime} \widetilde{\lambda}_{pq} V_p\rho^T V_q^\dagger \end{equation} 
restricting terms is not possible in the same way, as we can see in the following example. Define
\begin{eqnarray}
\ket{\Psi_\pm} &=&  \ket{0}\ket{0} \pm \ket{1}\ket{1}, \\
\ket{\Phi_\pm} &=&  \ket{0}\ket{1} \pm \ket{1}\ket{0} .
\end{eqnarray}
It is easy to verify that
\begin{equation}
\left( \oper{\Phi_-}{\Phi_-} \right)^{T_B} = \oper{0}{0} \otimes \oper{1}{1} + \oper{1}{1} \otimes \oper{0}{0} + \frac{1}{2} \left( \oper{\Psi_-}{\Psi_-} - \oper{\Psi_+}{\Psi_+} \right) \label{dec_e}
\end{equation}
and so
\begin{equation}
W=\oper{\Psi_+}{\Psi_+} + \frac{1}{2} \left( \oper{\Phi_-}{\Phi_-} \right)^{T_B} = \oper{0}{0} \otimes \oper{1}{1} + \oper{1}{1} \otimes \oper{0}{0} + \frac{1}{2} \oper{\Psi_-}{\Psi_-}.
\end{equation}
This kind of example shows that a similar restriction criterion to that given in Theorem \ref{id_test} is more difficult to establish in this more general case. It can be shown that the two subspaces  $\mathcal{W} = \{ \alpha\oper{\Phi_-}{\Phi_-} \ | \ \alpha \in \mathbb{R} \}$ and $\mathcal{W}^\prime = \{ \alpha\oper{\Psi_+}{\Psi_+} \ | \ \alpha \in \mathbb{R} \}$ satisfy the requirements of Theorem \ref{id_test} for $\mathcal{W}(\mathcal{V})$.\footnote{This fact is proved in the arguments contained in Theorem \ref{red_ind} for the first subspace; the property holds for the second subspace by a similar argument. We do not include a full proof here as we are simply trying to illustrate the added difficulty in restricting the decomposition of $W$ in this more general case.} However, in the decomposition $W=P+Q^{T_B}$, we can `mix up' the terms in some manner to get a different decomposition. It follows that we need some stronger conditions here to restrict the decomposition of $W$, and we do not attempt in this paper to give such extra conditions.

However, if a decomposable map $M: \mathcal{B}(\mathcal{H}) \to \mathcal{B}(\mathcal{H})$ exists such that 
\begin{equation} M(\Lambda(\rho)) = \sum_{m,n=1}^N \lambda_{mn} U_m\rho U_n^\dagger \end{equation}
then we can apply Theorem \ref{id_test} to this. If from this we can show that $M \circ \Lambda$ is indecomposable, then so is $\Lambda$, since if $\Lambda$ is decomposable, the decomposability of $M$ would imply that $M \circ \Lambda$ is decomposable. Similarly, the existence of a decomposable map $N: \mathcal{B}(\mathcal{H}) \to \mathcal{B}(\mathcal{H})$ such that
\begin{equation} N(\Lambda(\rho)) = \sum_{p,q=1}^{N^\prime} \widetilde{\lambda}_{pq} V_p\rho^T V_q^\dagger \end{equation}
could be used, via Theorem \ref{id_test}, to attempt to show that $\Lambda$ is not decomposable.

\section{Families of indecomposable maps} \label{sec_ex}

In this section we are going to present two families of positive maps that we can show are indecomposable. At first both families appear to have no common structure, but each map in both families can be shown to be indecomposable by the above theorem.

\subsection{The extended reduction criterion}

First, we will show how the reduction criterion \cite{Red1, Red2}, which arises from a decomposable map, can be improved upon by modifying the decomposable map so that it is still positive but no longer decomposable.\footnote{During the preparation of this manuscript, this construction also appeared in \cite{Br}, but is presented from a slightly different perspective.}

\subsubsection{The reduction criterion and its extension}

Let $\dim \mathcal{H}=d$, with a standard orthonormal basis $\{ \ket{k} \}_{k=0}^{d-1}$. The reduction map $R : \mathcal{B}(\mathcal{H}) \to \mathcal{B}(\mathcal{H})$ is defined by
\begin{equation} R(\sigma) = \Tr(\sigma) 1 - \sigma \end{equation}
where $\sigma \in \mathcal{B}(\mathcal{H})$. It can be shown that this is a positive but not a completely positive map, and so if $\rho \in \mathcal{B}(\mathcal{H}_1 \otimes \mathcal{H}_2)$ is a bipartite quantum state, then 
\begin{equation} (I \otimes R)(\rho) = \rho_1 \otimes 1_2 - \rho \geq 0, \end{equation}
(where $\rho_1 = \Tr_2 (\rho)$ is a reduced state of $\rho$, and $1_2$ is the identity matrix in $\mathcal{B}(\mathcal{H}_2)$) is a necessary condition for $\rho$ to be separable. This condition is known as the \emph{reduction criterion} \cite{Red1, Red2}. However, the reduction map is a completely co-positive map, and can be expressed as follows:
\begin{thm}[\cite{MRed}] \label{red_form} The reduction map $R: \mathcal{B}(\mathcal{H}) \to \mathcal{B}(\mathcal{H}) $ is decomposable and can be written as the form (for $\sigma \in \mathcal{B}(\mathcal{H})$): 
\begin{equation} R(\sigma) = \sum_{0\leq k<l\leq d-1} A_{kl} \sigma^T A_{kl}^\dagger \end{equation}
where $A_{kl} = \oper{k}{l} - \oper{l}{k}$. \end{thm}
and so the reduction criterion is weaker than the partial transpose condition.

Let $\sigma=\oper{\psi}{\psi}$ for some $\ket{\psi} \in \mathcal{H}$. Let $\{ \ket{\psi_k} \}_{k=1}^d$ be any orthonormal basis for $\mathcal{H}$, such that $\ket{\psi_1} \equiv \ket{\psi}$. Then we can write
\begin{equation} R(\sigma) = 1 - \oper{\psi}{\psi} = \sum_{k=2}^d \oper{\psi_k}{\psi_k} \label{red_pr} \end{equation}
i.e. we can write $R(\sigma)$ as the sum of $d-1$ orthogonal projections onto a state orthogonal to $\ket{\psi}$. Now suppose that we can find a positive linear map $S : \mathcal{B}(\mathcal{H}) \to \mathcal{B}(\mathcal{H})$ such that $S(\sigma) = \oper{\psi^\perp}{\psi^\perp}$, where $\inprod{\psi^\perp}{\psi} = 0$. Then, by letting $\ket{\psi_2}=\ket{\psi^\perp}$, we note that
\begin{equation} R(\sigma) - S(\sigma) = \sum_{k=3}^d \oper{\psi_k}{\psi_k} \end{equation}
i.e. we still have a positive linear map. Since both $R$ and $S$ are linear, this construction also works for $\sigma$ being a general mixed state.

The upshot of all this is that if we can construct a map $S$ as above, then we can construct a new positive map $R_E = R - S$, such that, if $\sigma \geq 0$, then $R_E(\sigma) \geq 0 \Leftarrow R(\sigma) \geq 0$. We will call a map $R_E$ constructed in this way an \emph{extended reduction map}. The next step of this construction is to explictly show how to construct such a map $S$.

Let us take a general normalised state $\ket{\psi} = \sum_{k=0}^{d-1} \alpha_k \ket{k}$. When $d=2$, there is (up to an overall phase) a unique state $\ket{\psi^\perp} = \alpha_1^*\ket{0} - \alpha_0^* \ket{1}$ orthogonal to $\ket{\psi} = \alpha_0 \ket{0} + \alpha_1 \ket{1}$. This state can be written as $\ket{\psi^\perp} = U\ket{\psi^*}$, where $U$ is the antisymmetric unitary operator $\oper{0}{1} - \oper{1}{0}$. We will try and generalise such an operation to a Hilbert space of dimension $d$

\begin{lemma} \label{asym} Let $\ket{\psi^\perp} = M\ket{\psi^*}$, where $M \in \mathcal{B}(\mathcal{H})$. Then $\inprod{\psi^\perp}{\psi} = 0$ for all $\ket{\psi}$ if and only if $M = -M^T$. \end{lemma}

\textbf{Proof} Let $M = \sum_{k,l} M_{kl} \oper{k}{l}$. Then 
\begin{equation}
\inprod{\psi^\perp}{\psi} = \sum_{k,l=0}^{d-1} M_{kl}^* \alpha_l \alpha_k = \sum_{k=0}^{d-1} M_{kk}^* \alpha_k^2 + \sum_{0 \leq k<l<d} (M_{kl}^*+M_{lk}^*) \alpha_l \alpha_k
\end{equation}
which is zero for all $\ket{\psi}$ if and only if the coefficients for each term is zero i.e. $M_{kl}=-M_{lk}$ for all $k,l=0,\ldots,d-1$, which is equivalent to $M=-M^T$. $\Box$

We should note that in Lemma \ref{asym}, we only required $M$ to be a linear operator. For our construction of the map $S$ however, we need that $\inprod{\psi^\perp}{\psi^\perp} = 1$ i.e. we also require $M$ to be \emph{unitary}. 

Putting all of this together, and making the observation that for Hermitian operators $\sigma$, $\sigma^T=\sigma^*$, we obtain the following theorem:

\begin{thm} Let $U \in \mathcal{B}(\mathcal{H})$ be unitary and antisymmetric, and let $R_E : \mathcal{B}(\mathcal{H}) \to \mathcal{B}(\mathcal{H})$ be a linear map defined by $R_E(\sigma) = \Tr(\sigma)1 - \sigma - U\sigma^TU^\dagger$. Then $R_E$ is a positive map.\footnote{Readers may have thought that it would be possible to subtract off two projectors from the right hand side of (\ref{red_pr}), so we end up with a map $R_E(\sigma) = \Tr(\sigma)1 - \sigma - U_1\sigma^TU_1^\dagger - U_2\sigma^TU_2^\dagger$, with $U_1, U_2$ unitary and antisymmetric. However for this to work we would need the terms subtracted off to be orthogonal i.e. we would need $\bra{\psi^*}U_2^\dagger U_1 \ket{\psi^*}$=0 for all $\ket{\psi}$, so $U_2^\dagger U_1=0$, which is impossible for unitary matrices because their rank is maximal. } \end{thm}

We note that antisymmetric unitaries only exist in even dimensions i.e.
\begin{lemma} Let $M \in \mathcal{B}(\mathcal{H})$ be antisymmetric, and $\dim \mathcal{H}=d$ be odd. Then $M$ cannot be unitary. \label{anti} \end{lemma}

\textbf{Proof} If $M$ is unitary, all of its eigenvalues are non-zero. However, the eigenvalues of an antisymmetric matrix $M$ occur in pairs $\pm \lambda$: Since $M$ is antisymmetric, it is normal and hence diagonalisable i.e. $M = \sum_i \lambda_i \oper{\psi_i}{\psi_i}$, where $\inprod{\psi_i}{\psi_j} = \delta_{ij}$. However, since $(\oper{\psi}{\psi})^T = \oper{\psi^*}{\psi^*}$, we also have $M = -M^T = \sum_i (-\lambda_i) \oper{\psi_i^*}{\psi_i^*}$ i.e. if $M\ket{\psi} = \lambda \ket{\psi}$, then $M\ket{\psi^*} = -\lambda \ket{\psi^*}$. Hence if $d$ is odd, $M$ must have a zero eigenvalue and hence cannot be unitary. $\Box$

When $d$ is even, we can construct antisymmetric unitaries easily. The unitary $U = VDV^T$ is antisymmetric, where $V \in U(d)$ is an arbitrary real orthogonal matrix, and $D$ is the antisymmetric unitary operator
\begin{equation} D = \sum_{k=0}^{d/2-1} e^{i\phi_k} \left( \oper{2k}{2k+1} - \oper{2k}{2k+1} \right) \end{equation}
where $\phi_k \in [0,2\pi ]$. The conjugation by $V$ on $D$ means that $U$ is unitary, and $U^T = VD^TV^T = -VDV^T = -U$ i.e. $U$ is antisymmetric. 
% Another construction of real antisymmetric unitaries comes from Hadamard matrices; for details see Appendix A.

\subsubsection{Indecomposability of the extended reduction map}

Here we use theorem \ref{id_test} to show that the extended reduction map is indecomposable:

\begin{thm} \label{red_ind} Let $\dim \mathcal{H}=d>2$ be even. The positive map $R_E: \mathcal{B}(\mathcal{H}) \to \mathcal{B}(\mathcal{H})$ defined by $R(\sigma) = \Tr(\sigma)1 - \sigma - U\sigma^TU^\dagger$, where $U$ is an antisymmetric unitary operator, is indecomposable. \end{thm}

\textbf{Proof} For $0 \leq k < l < d$, define $A_{kl}=\oper{k}{l} - \oper{l}{k}$. Since $U$ is antisymmetric, we can write $U = \sum_{k<l} U_{kl}A_{kl}$. Then, using theorem \ref{red_form}, we can rewrite $R_E(\sigma)$ in the form
\begin{eqnarray}
R_E(\sigma) &=& \sum_{0\leq k<l\leq d-1} A_{kl} \sigma^T A_{kl}^\dagger - \sum_{i<j, k<l} U_{ij}U_{kl}^*A_{ij}\sigma^T A_{kl}^\dagger \\
&=& \sum_{i<j, k<l} \Lambda_{(i,j),(k,l)} A_{ij}\sigma^T A_{kl}^\dagger 
\end{eqnarray}
where $\Lambda_{(i,j),(k,l)}$ is defined by
\begin{equation} \Lambda_{(i,j),(k,l)} = \left\{ \begin{array}{ll} 1 - |U_{ij}|^2 & (i,j) = (k,l) \\ -U_{ij}U_{kl}^* & (i,j) \neq (k,l) \end{array} \right. \end{equation}
Using the notation of Theorem \ref{id_test}, $\mathcal{V}$ is spanned by the set of antisymmetric matrices $\{ A_{kl} \}_{0 \leq k < l < d}$, and so $\mathcal{W}(\mathcal{V})$ has a basis $\{\ket{\Phi^-_{kl}} \equiv \ket{k}\ket{l} - \ket{l}\ket{k} \}_{0 \leq k < l < d}$. Furthermore, the matrix $L$ is a $d(d-1)/2 \times d(d-1)/2$ matrix defined by $\Lambda_{(i,j),(k,l)}$, where we consider the pairs $(i,j)$ and $(k,l)$ as the row and column indices. However, if we define the vector $\ket{U} = \sum_{i<j} U_{ij} \ket{(i,j)}$, then we have that $L=I-\oper{U}{U}$, and since $\inprod{U}{U} = \sum_{i<j} |U_{ij}|^2 = \frac{1}{2}\sum_{i,j} |U_{ij}|^2 = \frac{d}{2}$, $L$ has one negative eigenvalue $1-\frac{d}{2}$.
Hence to show $R_E$ is indecomposable, we must show that for all $Q \in \mathcal{B}(\mathcal{H})$, there exists $\ket{\psi} \in \mathcal{W}(\mathcal{V})^\perp$ such that $\bra{\psi}Q^{T_B}\ket{\psi}>0$. In this case, 
\begin{equation}
\mathcal{W}(\mathcal{V})^\perp = \textrm{span}(\{ \ket{k}\ket{k} \}_{0\leq k < d} \cup \{ \ket{\Phi^+_{kl}} \equiv \ket{k}\ket{l} + \ket{l}\ket{k} \}_{0 \leq k < l < d}). \end{equation}
First we note that, for $0\leq k<d$, 
\begin{equation}
\bra{k}\bra{k}Q^{T_B}\ket{k}\ket{k} = \bra{k}\bra{k}Q\ket{k}\ket{k} \geq 0.
\end{equation}
If $\bra{k}\bra{k}Q\ket{k}\ket{k}>0$, we are done. If not, then since $Q$ is positive, $\bra{k}\bra{k}Q\ket{l}\ket{l}=0$ for all $0\leq k,l<d$. Now define $\ket{\Psi^{\pm}_{kl}} = \ket{k}\ket{k} \pm \ket{l}\ket{l}$ for $0\leq k,l<d, k \neq l$. It is simple to verify that
\begin{equation}
\left( \oper{\Phi^+_{kl}}{\Phi^+_{kl}} \right)^{T_B} = \oper{k}{k} \otimes \oper{l}{l} + \oper{l}{l} \otimes \oper{k}{k} + \frac{1}{2} \left( \oper{\Psi^+_{kl}}{\Psi^+_{kl}} - \oper{\Psi^-_{kl}}{\Psi^-_{kl}} \right)
\end{equation} 
and hence
\begin{eqnarray}
\bra{\Phi^+_{kl}}Q^{T_B}\ket{\Phi^+_{kl}} &=& \Tr(Q^{T_B}\oper{\Phi^+_{kl}}{\Phi^+_{kl}} ) \\
&=& \Tr \left(Q \left( \oper{\Phi^+_{kl}}{\Phi^+_{kl}} \right)^{T_B} \right) \\
&=& \bra{k}\bra{l} Q \ket{k} \ket{l} + \bra{l}\bra{k} Q \ket{l} \ket{k} \geq 0 
\end{eqnarray}
where in the last expression the expectations of $Q$ with respect to $\ket{\Psi^\pm_{kl}}$ disappear because $\bra{k}\bra{k}Q\ket{l}\ket{l}=0$. If again all of these expectations are zero, this would imply that $\bra{k}\bra{l} Q \ket{k} \ket{l}=0$ for all $k,l$, and hence $\Tr(Q)=0$, implying $Q=0$. This completes the proof. $\Box$

\subsection{The positive maps of Piani}

In \cite{Piani}, a family of maps is proved to be positive. These maps are given by the following theorem:

\begin{thm}[\cite{Piani}] Let $\mathcal{H}_1, \mathcal{H}_2$ be Hilbert spaces of dimension $d_1, d_2$ respectively, and, for $k=1,2$, let $\{ F^{(k)}_\mu \}_{\mu=1}^{d_k^2}$ be Hermitian bases for $\mathcal{B}(\mathcal{H}_1), \mathcal{B}(\mathcal{H}_2)$ respectively satisfying $\Tr (F^{(k)}_\nu F^{(k)}_\mu) = \delta_{\mu \nu}$. Define $\Lambda_k : \mathcal{B}(\mathcal{H}_k) \to \mathcal{B}(\mathcal{H}_k)$ by 
\begin{equation} \Lambda_k(\rho) = \sum_{\mu=1}^{d_i^2} \lambda_\mu^{(k)} F^{(k)}_\mu \rho F^{(k)}_\mu \end{equation}
and $\Lambda: \mathcal{B}(\mathcal{H}_1 \otimes \mathcal{H}_2) \to \mathcal{B}(\mathcal{H}_1 \otimes \mathcal{H}_2)$
by $\Lambda = \Lambda_1 \otimes I_2 + I_1 \otimes \Lambda_2$. Then, if $\lambda_{d_2^2}^{(2)}<0$, and $\lambda_\mu^{(k)} \geq |\lambda_{d_2^2}^{(2)}|$ for all $\mu$ when $k=1$ and all $\mu \neq d_2^2$ when $k=2$, then $\Lambda$ is a positive map. \end{thm}

In \cite{Piani}, bound entangled states were constructed to show a subset of these maps were indecomposable. However, Theorem \ref{id_test} can be used to show that a much larger class of these these maps are indecomposable.

\begin{thm} In the above definition, let $F^{(k)}_1 = I_k$ for $k=1,2$ respectively. Then $\Lambda$ defined as above is indecomposable. \end{thm}

\textbf{Proof} Define $\mathcal{H}= \mathcal{H}_1 \otimes \mathcal{H}_2$, with standard orthonormal basis $\{ \ket{k} \}_{k=0}^{d_1-1}$, $\{ \ket{l} \}_{l=0}^{d_2-1}$ respectively. For $\rho \in \mathcal{B}(\mathcal{H})$, we may write
\begin{eqnarray}
\Lambda(\rho) &=& \sum_{\mu=1}^{d_1^2} \lambda_\mu^{(1)} (F^{(1)}_\mu \otimes I_2) \rho \left( F^{(1)}_\mu \otimes I_2 \right)^\dagger + \sum_{\mu=1}^{d_2^2} \lambda_\mu^{(2)} (I_1 \otimes F^{(2)}_\mu) \rho (I_1 \otimes F^{(2)}_\mu)^\dagger \\
&=& (\lambda^{(1)}_1+\lambda^{(2)}_1) \rho + \sum_{\mu=2}^{d_1^2} \lambda_\mu^{(1)} (F^{(1)}_\mu \otimes I_2) \rho \left( F^{(1)}_\mu \otimes I_2 \right)^\dagger \nonumber \\
&& \hspace{4cm} + \sum_{\mu=2}^{d_2^2} \lambda_\mu^{(2)} (I_1 \otimes F^{(2)}_\mu) \rho (I_1 \otimes F^{(2)}_\mu)^\dagger
\end{eqnarray}

Using the notation of Theorem \ref{id_test}, the subspace $\mathcal{V}$ has a basis $\{ I_1 \otimes I_2 \} \cup \{ F^{(1)}_\mu \otimes I_2 \}_{\mu=2}^{d_1^2} \cup \{ I_1 \otimes F^{(2)}_\mu \}_{\mu=2}^{d_2^2}$. Hence 
\begin{equation} \mathcal{V} = \{ M_1 \otimes I_2 + I_1 \otimes M_2 \ | \ M_i \in \mathcal{B}(\mathcal{H}_i); i=1,2 \}. \end{equation} 
Furthermore, $L=\textrm{diag}(\lambda^{(1)}_1+\lambda^{(2)}_1,\lambda^{(1)}_2, \ldots,\lambda^{(1)}_{d_1^2},\lambda^{(2)}_2,\ldots,\lambda^{(2)}_{d_2^2})$, and so $L$ is not positive. 

Let us now consider $\mathcal{W}(\mathcal{V})$. First, define $\ket{\Psi(d_i)^+} = \sum_{m=0}^{d_i-1} \ket{m}\ket{m} \in \mathcal{H}_i \otimes \mathcal{H}_i$ (the maximally entangled state). For notational convenience, we will rearrange the tensor product $\mathcal{H}_1 \otimes \mathcal{H}_2 \otimes \mathcal{H}_1 \otimes \mathcal{H}_2$ into the order $\mathcal{H}_1 \otimes \mathcal{H}_1 \otimes \mathcal{H}_2 \otimes \mathcal{H}_2$ (as we may). The subspace $\mathcal{W}(\mathcal{V}) \in \mathcal{B}(\mathcal{H}_1 \otimes \mathcal{H}_1 \otimes \mathcal{H}_2 \otimes \mathcal{H}_2)$
is a $d_1^2 +d_2^2 -1$ dimensional subspace with a basis consisting the vectors
\begin{enumerate}
\item The vector $\ket{\Psi(d_1)^+}\ket{\Psi(d_2)^+}$;
\item A basis of $d_1^2-1$ vectors in the subspace $\{ \ket{\Psi(d_1)^+}\ket{\Psi_2^\perp} \ | \ \inprod{\Psi_2^\perp}{\Psi(d_2)^+}=0\}$;
\item A basis of $d_2^2-1$ vectors in the subspace $\{ \ket{\Psi_1^\perp}\ket{\Psi(d_2)^+} \ | \ \inprod{\Psi_1^\perp}{\Psi(d_1)^+}=0\}$.
\end{enumerate}
Hence we can write
\begin{equation}
\mathcal{W}(\mathcal{V})^\perp = \textrm{span}\left( \{ \ket{\Psi_1^\perp}\ket{\Psi_2^\perp} \ | \ \inprod{\Psi_i^\perp}{\Psi(d_i)^+}=0, i=1,2 \}\right).
\end{equation}

In the tensor product $\mathcal{H}_1 \otimes \mathcal{H}_1 \otimes \mathcal{H}_2 \otimes \mathcal{H}_2$, let us relabel the indices $A1,B1,A2,B2$ from left to right. Then, from Theorem 2 all that remains to be shown is that for all $Q \in \mathcal{B}(\mathcal{H}_{A1} \otimes \mathcal{H}_{B1} \otimes \mathcal{H}_{A2} \otimes \mathcal{H}_{B2})$, there exists $\ket{\psi} \in \mathcal{W}(\mathcal{V})^\perp$ such that $\bra{\psi}Q^{T_B}\ket{\psi}>0$, where $T_B$ now represents the partial transposition of systems $B1$ and $B2$ together. From here on in, we will place indices on bra and kets to indicate which Hilbert space each belongs to.

First we observe that if (for $i=1,2$), $0\leq k_i,l_i <d_i; k_i \neq l_i$, then $\ket{k_1}_{A1}\ket{l_1}_{B1}\ket{k_2}_{A2}\ket{l_2}_{B2} \in \mathcal{W}(\mathcal{V})^\perp$. We also note that
\begin{eqnarray}
\bra{k_1}_{A1}\bra{l_1}_{B1}\bra{k_2}_{A2}\bra{l_2}_{B2}Q^{T_B}\ket{k_1}_{A1}\ket{l_1}_{B1}\ket{k_2}_{A2}\ket{l_2}_{B2} &=& \nonumber \\
\bra{k_1}_{A1}\bra{l_1}_{B1}\bra{k_2}_{A2}\bra{l_2}_{B2} Q \ket{k_1}_{A1}\ket{l_1}_{B1}\ket{k_2}_{A2}\ket{l_2}_{B2} &\geq& 0. \label{e1}
\end{eqnarray}
If any of these quantities are positive then we are done. If all of these expectations are zero, then since $Q$ is positive, necessarily 
\begin{equation}
\bra{k_1}_{A1}\bra{l_1}_{B1}\bra{k_2}_{A2}\bra{l_2}_{B2} Q \ket{k^\prime_1}_{A1}\ket{l^\prime_1}_{B1}\ket{k^\prime_2}_{A2}\ket{l^\prime_2}_{B2}=0 \label{q_pos}
\end{equation}
(where for $i=1,2$, $0\leq k^\prime_i,l^\prime_i <d_i; k^\prime_i \neq l^\prime_i$). 
Furthermore, let us define 
\begin{equation} \ket{\Phi^\perp_i} = \sum_{m=0}^{d_i-1} \omega_{d_i}^m \ket{m}_{Ai} \ket{m}_{Bi} \in \mathcal{H}_i \otimes \mathcal{H}_i \end{equation}
for $i=1,2$, where $\omega_d = \exp( 2 \pi i/d)$. It is easy to verify that $\inprod{\Phi^\perp_i}{\Psi^+(d_i)} = 0$, and hence $\ket{\Phi^\perp_1}\ket{\Phi^\perp_2} \in \mathcal{W}(\mathcal{V})^\perp$. Furthermore,
\begin{equation}
\oper{\Phi^\perp_i}{\Phi^\perp_i}^{T_{Bi}} = \sum_{m=0}^{d_i-1} \oper{m}{m}_{Ai} \otimes \oper{m}{m}_{Bi} + P_i 
\end{equation}
where $P_i$ has support on the subspace spanned by $\{\ket{k}_{Ai}\ket{l}_{Bi} \}_{0 \leq k,l < d; k \neq l}$. Hence 
\begin{eqnarray}
&\bra{\Phi^\perp_1}\bra{k_2}_{A2}\bra{l_2}_{B2}Q^{T_B}\ket{\Phi^\perp_1}\ket{k_2}_{A2}\ket{l_2}_{B2}& \\
=& \Tr \left( Q^{T_B} \left( \oper{\Phi^\perp_1}{\Phi^\perp_1} \otimes \oper{k_2}{k_2}_{A2} \otimes \oper{l_2}{l_2}_{B2} \right) \right) & \\
=& \Tr \left( Q \left( \oper{\Phi^\perp_1}{\Phi^\perp_1} \otimes \oper{k_2}{k_2}_{A2} \otimes \oper{l_2}{l_2}_{B2} \right)^{T_B} \right) & \\
=& \sum_{m=0}^{d_1-1} \bra{m}_{A1}\bra{m}_{B1}\bra{k_2}_{A2}\bra{l_2}_{B2} Q \ket{m}_{A1}\ket{m}_{B1}\ket{k_2}_{A2}\ket{l_2}_{B2} & \geq 0 \label{e2}
\end{eqnarray}
where any expectations from the $P_1$ term disappear because of (\ref{q_pos}). Similarly,
\begin{eqnarray}
&\bra{k_1}_{A1}\bra{l_1}_{B1}\bra{\Phi^\perp_2}Q^{T_B}\bra{k_1}_{A1}\bra{l_1}_{B1}\ket{\Phi^\perp_2}& \\
=& \sum_{n=0}^{d_2-1} \bra{k_1}_{A1}\bra{l_1}_{B1}\bra{n}_{A2}\bra{n}_{B2} Q \ket{k_1}_{A1}\ket{l_1}_{B1}\ket{n}_{A2}\ket{n}_{B2} &\geq 0. \label{e3}
\end{eqnarray}
Again, if either of these expectations is positive, we are done. If both are zero, then again positivity implies that
\begin{eqnarray} 
\bra{m}_{A1}\bra{m}_{B1}\bra{k_2}_{A2}\bra{l_2}_{B2}Q\ket{m^\prime}_{A1}\ket{m^\prime}_{B1}\ket{k^\prime_2}_{A2}\ket{l^\prime_2}_{B2}&=&0; \\
\bra{k_1}_{A1}\bra{l_1}_{B1}\bra{n}_{A2}\bra{n}_{B2}Q\ket{k^\prime_1}_{A1}\ket{l^\prime_1}_{B1}\ket{n^\prime}_{A2}\ket{n^\prime}_{B2}&=&0,
\end{eqnarray}
with $0\leq m,m^\prime <d_1, 0\leq n,n^\prime < d_2$. Finally, with these two expressions we see that
\begin{eqnarray}
&\bra{\Phi^\perp_1}\bra{\Phi^\perp_2}Q^{T_B}\ket{\Phi^\perp_1}\ket{\Phi^\perp_2}& \\
=& \sum_{m=0}^{d_1-1} \sum_{n=0}^{d_2-1} \bra{m}_{A1}\bra{m}_{B1}\bra{n}_{A2}\bra{n}_{B2} Q \ket{m}_{A1}\ket{m}_{B1}\ket{n}_{A2}\ket{n}_{B2} &\geq 0. \label{e4}
\end{eqnarray}

If this expectation is zero, then equations (\ref{e1}), (\ref{e2}), (\ref{e3}) and (\ref{e4}) combined imply 
\begin{equation}
\bra{k_1}_{A1}\bra{l_1}_{B1}\bra{k_2}_{A2}\bra{l_2}_{B2} Q \ket{k_1}_{A1}\ket{l_1}_{B1}\ket{k_2}_{A2}\ket{l_2}_{B2} = 0
\end{equation}
for all $0\leq k_i,l_i <d_i; i=1,2$, which in turn implies $\Tr(Q)=0$, which (since $Q$ is positive) would imply $Q=0$. This completes our proof. $\Box$

\section{Conclusion} \label{sec_con}

In this paper we have given a sufficient criterion for a given positive map to be indecomposable, and we have used it to verify a number of positive maps are not decomposable, as well as constructing new examples. While this condition does not immediately give a method of constructing positive indecomposable maps, it is a first step to examining the structure of these maps.

The conditions we placed on $\mathcal{W}(\mathcal{V})$ were at least as strong as enforcing this subspace to contain no operators with a positive partial transpose i.e. we might have expected that it would be sufficient to enforce all operators in $\mathcal{W}(\mathcal{V})$ to have a negative eigenvalue under partial transposition for our criterion. While we have been unable to prove this would be a sufficient condition to restrict the form of the entanglement witness as required, we have not disproved it either, and so this could be an issue for further exploration.

The first given example of a non-decomposable map was that given by Choi \cite{Ch2}. This map $C: \mathcal{B}(\mathbb{C}^3) \to \mathcal{B}(\mathbb{C}^3)$ can be written in the form
\begin{equation} \sum_{k=0}^2 \left(2 P_{kk}\rho P_{kk} + 2 P_{k-1 k}\rho P_{k-1 k} \right) - \rho \end{equation}
where $P_{ij} = \oper{i}{j}$, and we are assuming modulo 3 addition in the indices of these projectors in the above expression.
Unfortunately we cannot seem to prove this map is not decomposable by the criterion outlined in this paper. Since many of the examples of indecomposable maps in the literature are generalisations of this map \cite{ChGen, ChGen2, ChGen3, ChGen4, ChGen5, ChGen6}, it would be a sensible next step to see if the condition above can be generalised in any way to include this important class of positive maps. An even more ambitious but very useful step would be to see if this work can be used to characterise some of the structure that may be present in a non-decomposable map.

\subsection*{Acknowledgments}
The author would like to thank Tony Sudbery for reading this manuscript and for his continuing support, and the Engineering and Physical Sciences Research Council (U.K.) for funding this research.

\end{document}